\documentclass{ar-1col-S2O}

\usepackage[numbers]{natbib}
\usepackage{url}
\usepackage{graphicx}
\usepackage{enumitem}
\usepackage{mathtools}
\usepackage{graphicx}   
\usepackage{color}
\usepackage[normalem]{ulem}
\usepackage{caption}
\usepackage{subcaption}
\usepackage{braket}
\usepackage{comment}
\usepackage{bm}
\usepackage{array}
\usepackage{booktabs}
\usepackage{multirow}

\usepackage{amsmath}

\usepackage{booktabs}

\usepackage[breaklinks=true]{hyperref}
\usepackage{breakcites}
\usepackage{algorithm}
\usepackage{algpseudocode}
\usepackage{xcolor}
\usepackage{bm}

\usepackage[]{draftwatermark}
\SetWatermarkScale{4}

\begin{document}
\title{Enhanced Sampling with Machine Learning: A Review}

\author{Shams Mehdi,$^{1,2}$ Zachary Smith,$^{1,2}$ Lukas Herron,$^{1,2}$ Ziyue Zou,$^3$ and Pratyush Tiwary$^{1,3}$
\affil{$^1$Institute for Physical Science and Technology,
 University of Maryland, College Park 20742, USA; email: ptiwary@umd.edu}
 \affil{$^2$Biophysics Program, 
 University of Maryland, College Park 20742, USA}
 \affil{$^3$Department of Chemistry and Biochemistry,
 University of Maryland, College Park 20742, USA}}

\begin{keywords}
Molecular dynamics, enhanced sampling, machine learning, artificial neural networks
\end{keywords}

	\date{\today}
\begin{abstract}
Molecular dynamics (MD) enables the study of physical systems with excellent spatiotemporal resolution but suffers from severe time-scale limitations. To address this, enhanced sampling methods have been developed to improve exploration of configurational space. However, implementing these is challenging and requires domain expertise. In recent years, integration of machine learning (ML) techniques in different domains has shown promise, prompting their adoption in enhanced sampling as well. Although ML is often employed in various fields primarily due to its data-driven nature, its integration with enhanced sampling is more natural with many common underlying synergies. This review explores the merging of ML and enhanced MD by presenting different shared viewpoints. It offers a comprehensive overview of this rapidly evolving field, which can be difficult to stay updated on. We highlight successful strategies like dimensionality reduction, reinforcement learning, and flow-based methods. Finally, we discuss open problems at the exciting ML-enhanced MD interface.
\end{abstract}

	\maketitle
 \tableofcontents
\section{INTRODUCTION}
\label{sec:Introduction}

Molecular dynamics (MD) simulations play a crucial role in the field of physical chemistry and allied sciences, offering a powerful tool to investigate the intricate motions and behaviors of atoms and molecules. These simulations act as a virtual microscope, allowing scientists to explore the dynamic aspects underlying complicated processes. MD is implemented by discretizing time into small steps and Newton's equations of motion serve as the guiding principle for iterative generation of the time evolution of a system from initial atomic coordinates \cite{frenkel2000molecular}. This approach enables the study of the microscopic state of a system described by the position and momentum of each atom in phase space. In addition to the deterministic forces described by Newton's laws, MD simulations incorporate thermostats \cite{vrescale} to sample the canonical, constant number, volume, temperature (NVT) ensemble  or both thermostats and barostats \cite{PR_barostat} to sample the isothermal-isobaric, constant number, pressure, temperature (NPT)  ensemble. These techniques enhance the simulation's ability to reproduce realistic conditions and achieve accurate results.

Under this framework, the interactions between different entities present in a physical system are defined using numerical constants known as force fields that are obtained empirically or from first principle calculations. Depending on the task at hand, force fields with different levels of detail e.g., quantum-mechanical, classical, coarse-grained, etc. can be employed. In particular, classical force fields are obtained by carefully parametrizing atomic interactions to reproduce equilibrium properties observed in experimental studies \cite{hollingsworth2018molecular}. While these simulations excel in capturing equilibrium properties over long periods, sampling rare events becomes a challenging task when the integration step is on the order of femtoseconds ($fs$). A small $fs$ time step is necessary because in classical force fields, the fastest motion i.e., vibrations of hydrogen atoms take place at a similar time scale \cite{hollingsworth2018molecular}.

However, practical processes of interest such as large conformational shifts in proteins \cite{lindorff2011fast} or ligand binding/unbinding events \cite{dasat_koff, tiwary2015kinetics, shekhar2022protein} can occur on timescales ranging from milliseconds to hours. Even slower but critically important nucleation processes \cite{tsai2019reaction, zou2021toward, zou2023driving} may span seconds to days. Capturing these events using standard MD simulations can be computationally demanding, requiring an enormous amount of time. In fact, sampling a single event within these timescales may necessitate millions of years of computational effort. Although hardware advancements \cite{GPU-MD, Anton} have facilitated faster MD simulations, achieving the exponential speedup necessary to access these rare events remains a significant challenge due to the inherently sequential nature of time. Evidently even on the best available hardware such as the Anton supercomputer, sampling of rare events remains a challenge. Thus, researchers continue to explore alternative creative methods and algorithms to overcome these limitations and efficiently explore the dynamics of complex processes.

\begin{figure}
    \centering
    \includegraphics[
    width=\textwidth
    ]{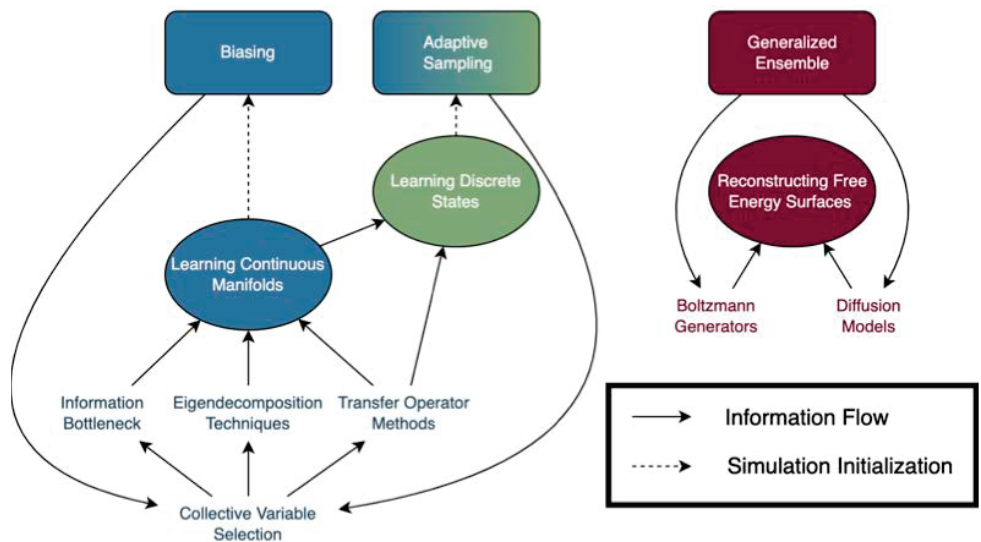}
    \caption{Overview of enhanced sampling methods and their interactions with ML methods. Enhanced sampling methods are shown as rectangles with colors corresponding to the associated learning tasks which are shown as ovals. Individual ML methods are shown as text with arrows corresponding to the information flow between methods. Note that ML for generalized ensemble methods is purely post-processing while learning for biasing and adaptive sampling methods informs new simulations.}
    \label{fig:summary}
\end{figure}

Enhanced sampling methods aim to address this issue by increasing the efficiency of exploring the configuration space and accelerating MD simulations. These techniques help to overcome high energy barriers and explore system states that are typically inaccessible in conventional simulations, potentially improving the accuracy of calculated thermodynamic and kinetic properties. However, the implementation of enhanced sampling algorithms is not trivial and may require significant human expertise.

In recent years, machine learning (ML) models have been employed in various domains, for example in the prediction of binding affinities of protein targeting small molecules for drug discovery \cite{s2017supervised}, in genomics and proteomics data analysis \cite{olson2018data}, synthesizing novel materials with tailored properties \cite{wei2019machine} etc. These
successful applications of ML in diverse scientific fields have inspired the adoption of similar
techniques in accelerating MD simulations. In this review, we examine the latest advancements in the field of ML-augmented enhanced sampling methods. Specifically, we will concentrate on enhancing the sampling capabilities of classical MD simulations. For readers interested in the application of ML in accelerating coarse-grained simulations, we refer to recent literature sources \cite{cisneros2016modeling,jin2022bottom,wang2021multi}. It is important to note that our focus will be solely on the utilization of ML to expedite simulations and not on the analysis of MD data \cite{wang2020machine,noe2020machine,rydzewski2023manifold}.

While often ML is applied to different fields solely due to the possibility of a data-driven approach, its confluence with enhanced sampling (Section \ref{sec:EnhancedSampling}) can be more organic. Under different names, to some extent, both disciplines have tackled similar problems. This could be the problem of dimensionality reduction (Section \ref{sec:dim_red}),  new strategies for improved bias deposition (Section \ref{sec:LearningEnhance}), or the problem of moving back-and-forth between tractable and intractable probability distributions (Section \ref{sec:flow}). This review looks at ML and enhanced MD through these and other shared lenses, summarizing the state-of-the-art in a burgeoning field that is hard to keep up with.
 
\section{ENHANCED SAMPLING}
\label{sec:EnhancedSampling}

Many enhanced sampling methods have been developed to tackle the timescale problem with molecular dynamics. We classify these methods into three distinct classes with different mechanisms and opportunities for synergies with ML. However, note that alternative classification schemes for enhanced sampling methods exist in the literature \cite{valsson2016enhancing,livecoms, prxUnified}. Our three classes are biasing methods, adaptive sampling methods, and generalized ensemble methods.

\begin{marginnote}
\entry{Collective variable (CV)}{Functions of input coordinates that offer a simplified description of the system's structure, though not necessarily any mechanistic insight about slow modes. CVs are typically smaller in number compared to the total degrees of freedom. For example, certain protein folding processes at physiological temperature can be represented by only three CVs \cite{piana2008advillin,altis2007dihedral}.}
\end{marginnote}

Biasing methods perform importance sampling by modifying the simulation with a bias potential that can be reweighted to recover unbiased statistics \cite{PTReweight, MBReweight}. This potential can be static \cite{torrie1977monte} or updated over the course of the simulation \cite{valsson2016enhancing, OPES} and is defined in terms of a small number of  collective variables (CVs) and not the full configuration of the system. These CVs can be simple basis functions such as distances or dihedral angles or they can be more complex linear or nonlinear combinations of basis functions. Determining which CVs to use in a data-driven manner can be considered a manifold learning problem where the goal is to find a low-dimensional manifold that effectively describes the system's relevant slow dynamics and/or dominant metastable states.

Adaptive sampling methods, also known as path sampling methods \cite{bolhuis2002transition,chong2017path}, perform importance sampling by strategically initializing rounds of short parallel simulations in states that are either under-sampled or likely to sample an unexplored state. They are often analyzed by constructing a Markov state model (MSM) \cite{MSMbook, MSMperspective} to combine the statistics and kinetics of these simulations. The separation of states can be done using geometric criteria, kinetic criteria, or even by discretizing CVs \cite{perez2013identification, pandeTica}. Initial quantitative comparisons between strategies have shown that different techniques are beneficial for exploring state space or sampling rare events and that \textit{a priori} knowledge can be used to improve sampling further \cite{CCComparison, GBComparison}. Adaptive sampling provides many opportunities for ML because states can be defined by either learning a continuous manifold or a more direct mapping from configurations to discrete states.

Generalized ensemble methods accelerate sampling by allowing the simulation to transition to a different ensemble with a different temperature, pressure, or Hamiltonian. These ensembles can have lower kinetic barriers between configurations and new free energy minima can be sampled in the original ensemble after crossing lower barriers in another ensemble. For example, transitioning to a higher temperature ensemble would accelerate barrier crossing, then transitioning to the original ensemble would allow sampling of a new energy minimum at the temperature of interest. This is often done with replicas occupying a ladder of ensembles and periodically exchanging ensembles in the case of replica exchange but can also be done with a single simulation in expanded ensemble methods \cite{ExpandedEnsembles, SimulatedTempering, SimulatedScaling}. This class of enhanced sampling methods provides different opportunities for ML as there is no requirement for a learned representation of the system's states. Instead, ML is used to analyze these simulations and to infer free energy surfaces, potentially for regions only sampled in some ensembles, with sampling from the other ensembles.

A number of methods such as replica exchange umbrella sampling \cite{sugita2000multidimensional}, parallel tempering metadynamics \cite{bussi2006free}, and bias exchange metadynamics \cite{piana2007bias} combine multiple of these classes at the same time. A longer list of hybrid methods with an elegant taxonomy can be found in Ref. \citenum{livecoms}.

\section{DIMENSIONALITY REDUCTION FOR SAMPLING}
\label{sec:dim_red}

\begin{figure}
    \centering
    \includegraphics[
    width=\textwidth
    ]{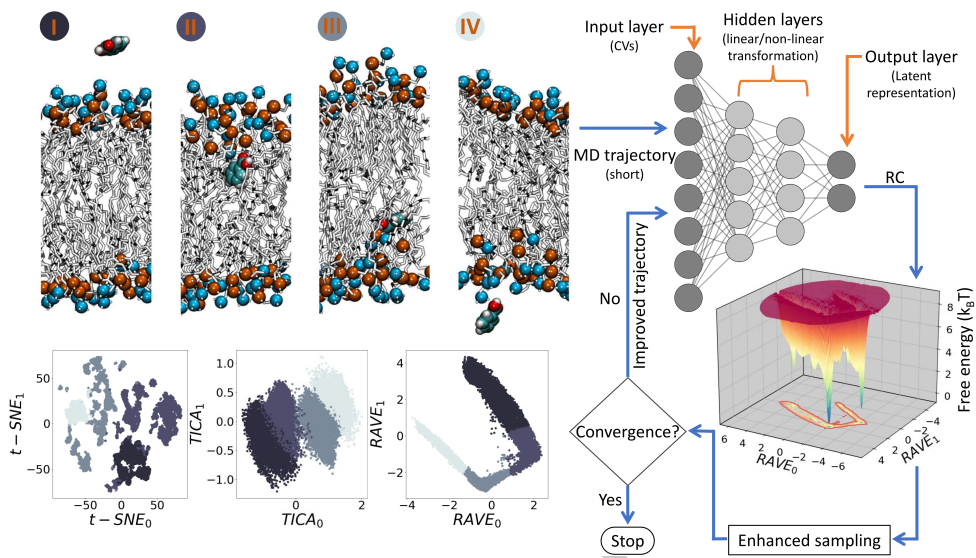}
    \caption{Small molecule permeation through a lipid bilayer. The general purpose dimensionality reduction method (t-SNE) fails to preserve kinetic information, but methods designed for MD data (TICA, RAVE) were successful. Permeation involves the sequential movement of the small molecule from one side (I), across the membrane (II, III), to the other side (III). A general protocol for iterative learning of improved RC is shown on the right panel.}
    \label{fig:dim_red_fig}
\end{figure}

In recent years, a dominant area of research in accelerating MD simulations through ML has been the development of dimensionality reduction techniques for identifying slow modes from simulated trajectories. Constraints to the atomic degrees of freedom in molecular systems generate this low-dimensional manifold, the study of which is driven by the widespread use of enhanced sampling methods e.g, umbrella sampling, metadynamics, weighted ensemble, milestoning, variationally enhanced sampling (VES) and others \cite{tiwary2013accelerated,tiwary2013metadynamics,zuckerman2017weighted,west2007extending,torrie1977monte},  which still, however, require \textit{a priori} identification of approximate Reaction Coordinates (RCs) describing system's slow degrees of freedom. By employing these methods and enhancing sampling along an approximate RC, rare events of interest can be observed. However, determining RCs for practical systems is typically challenging as they are often unknown \textit{a priori} and are difficult to identify without simulating the rare event of interest itself. ML based methods attempt to solve this problem by projecting the high-dimensional MD data from an arbitrarily long simulation onto a low-dimensional manifold designed to approximate the system's RC, often as a combination of a much bigger dictionary of CVs. Since the initial short simulation will typically not include the rare event of interest, the quality of RC can be improved by iterating between performing enhanced sampling and learning better RCs until convergence as illustrated in \textbf{Figure \ref{fig:dim_red_fig}}. In this context, it is important to note that there is a zoo of dimensionality reduction algorithms already available in other scientific domains. However, these methods are not always directly suitable for MD simulations because they are not designed to preserve kinetic information and fail to capture essential physics governing system behavior. This is illustrated in \textbf{Figure \ref{fig:dim_red_fig}}, where MD simulation data describing the permeation of a small molecule through a lipid bilayer \cite{mehdi2022accelerating} has been analyzed using a general purpose method (t-SNE \cite{van2008visualizing}), and methods developed for identifying approximate RC for MD (TICA \cite{m2017tica}, RAVE \cite{ribeiro2018reweighted}) respectively. It can be clearly observed that TICA and RAVE were able to correctly preserve kinetic information in addition to distinguishing the metastable states, which t-SNE failed to do.

\begin{marginnote}
\entry{Reaction coordinate (RC)}{An optimal coordinate representing progress along a reaction pathway that offers mechanistic insight. RCs are typically constructed as a combination of input CVs.}
\entry{Loss}{A measure of error or discrepancy between the predicted output of a ML model and expected output.}
\entry{Objective function}{A mathematical function that guides the search for the optimal solution by iteratively adjusting model parameters until a satisfactory solution is achieved.}
\end{marginnote}

In this section, we will consider data-driven dimensionality reduction methods for MD which typically involves the construction of an artificial neural network (ANN) and minimizing the loss of a well-defined objective function to generate a regularized low-dimensional manifold, also known as the latent space. In general, these recently developed methods draw inspiration from diverse approaches, which often overlap with each other and a meaningful classification of these approaches becomes a difficult task. In the following subsections, we attempt to classify them by primarily looking at the theoretical foundations of the adopted objective functions, and the specific purpose behind dimensionality reduction e.g., expressivity and interpretability. Additionally, ML methods such as the aforementioned vanilla t-SNE that could be used for clustering and analyzing MD trajectories, that are not necessarily suitable for enhanced sampling, are outside the scope of this review. Interested readers can refer to previous literature reviews that cover this specific topic \cite{wang2020machine}.

\subsection{AUTOMATED CV SELECTION}
\label{subsec:CV_sel}
Instead of constructing an abstract low-dimensional latent space or RC directly as a function of the entire input feature space, ML methods can be employed to identify the subset of CVs most complete for describing the system's behavior. For example, in a seminal  work by Dinner \textit{et al.} \cite{ma2005automatic} a genetic neural network algorithm was implemented to acquire the initial set of coordinates
that can clearly and effectively determine the transition state for a simple biomolecular transition. The authors showed how they could reproduce the committor from a set of CVs. The set of CVs that best represented the correct committor was then chosen to obtain the umbrella potential term, $V = k(p_B^{GNN}-p_B^{target})^2$ for enhanced sampling.

\begin{marginnote}
\entry{Committor}{In a chemical reaction involving two distinct states, the committor quantifies the probability or fraction of trajectories starting in the reactant state that will eventually reach the product state.}
\end{marginnote}

In a later, also pioneering work by Peters \textit{et al.} \cite{peters2006obtaining}, sets of potential CVs were examined for the construction of an appropriate RC using likelihood maximization under the transition path sampling (TPS) scheme. For an efficient screening of the CVs, a modified version of TPS called the aimless shooting algorithm was adopted to remove momentum correlations across each TPS trial trajectory. Finally, a good RC is identified by analyzing shooting history across trial trajectories and adopting a Bayesian information criterion for discarding complex models. Mathematically, if $p(TP|\textbf{x})$ represents the probability that the system will adopt a particular transition path given shooting point $\textbf{x}$, and $r(\textbf{x})$ represents an appropriate RC, then this method calculates $p(TP|r(\textbf{x}))$ corresponding to $p(TP|\textbf{x})$.

Both of these influential works considered the question of not just constructing a set of complete CVs, but also building a RC from them. In a recent study \cite{ravindra2020automatic}, Ravindra \textit{et al.} introduced a method for only the first part of the problem, i.e. CV identification, called Automatic Mutual Information Noise Omission (AMINO) for the automated selection of CVs from MD data. Although AMINO does not directly create RCs for enhanced sampling, it generates a subset of the most relevant CVs describing a system by discarding correlated, or noisy CVs. This subset of CVs can be used  to calculate improved RCs through other RC construction approaches discussed in this review. AMINO operates by initially employing a mutual information-based distance metric ($D$) to determine the similarity between pairs of CVs ($X$ and $Y$).
\begin{equation}
\label{eq_pande}
D(X;Y) = 1- \frac{I(X;Y)}{H(X,y)}
\end{equation}
Here, the term $I$ represents the mutual information between $X$ and $Y$, while $H$ represents the joint entropy of $X$ and $Y$. This measure of similarity ($D$) is then utilized to group all the CVs into distinct clusters using K-medoids clustering. Subsequently, a single CV from each cluster is chosen for representing the CVs within the corresponding cluster as best as possible. Finally, the optimal number of CVs describing a dataset is determined by employing the jump method from rate-distortion theory. This involves constructing a distortion function and selecting the number of CVs that yields the greatest reduction in the distortion function.
\begin{marginnote}
\entry{Distortion function}{Mathematical function describing information loss in using a subset of CVs compared to all the available CVs describing a dataset.}
\end{marginnote}

In a subsequent study, Stock \textit{et al.} \cite{diez2022correlation} extended the idea of using mutual information as a similarity measure for selecting relevant CVs. Here, different clustering algorithms were explored and the authors found that the Leiden algorithm from graph theory for detecting communities of similar coordinates worked really well. The key idea is to examine the underlying graph structure constructed from system CVs as nodes, and mutual information-based similarity as edges. Leiden algorithm uses the definition of modularity in community detection for maximizing the following objective function ($\Phi$):
\begin{equation}
\label{eq_pande}
\Phi = \frac{1}{2m}\sum_c(e_c-\frac{k_c^2}{2m})
\end{equation}
In this equation, the subscript $c$ represents different clusters, while $m$, $e_c$, $k_c$ represent the number of edges, sum of edge weights, and sum of CV degrees within each cluster respectively. By applying this approach, the study aimed to identify communities or groups of CVs that exhibit higher similarity within each group compared to other groups.

Very recently, Ensing \textit{et al.} \cite{hooft2021discovering} expanded on the genetic algorithm-based approach introduced by Ma \& Dinner \cite{ma2005automatic} discussed earlier in this section. Here, the key idea is to employ a feed-forward ANN that takes subsets of CVs selected using a genetic algorithm as inputs and predicts atomic coordinates as the output. After training a ML model, fitness scores are assigned to each CV determined from the mean absolute error between ANN output and the ground truth atomic coordinates. A second genetic algorithm is incorporated for tuning the construction of the ANN architecture. In contrast to the approach taken by Dinner \textit{et al.}, this method bypasses the costly committor calculations by implementing a TPS framework for generating training data. Predicting full atomic coordinates as the model output enables the generation of configurations for unexplored regions which can help with initiating additional simulations. Lastly, a lag time between input and predicted atomic coordinates in the output can be introduced to identify CVs appropriate for determining the slow modes.

\subsection{TRANSFER OPERATOR APPROXIMATION}
\label{subsec:transfer_operator}
One popular approach for reducing the dimensionality of MD trajectories through ML has been the identification of the slowest eigenfunctions $\psi_i(\textbf{x})$ of a system's transfer operator ($\mathcal{T}$). Assuming the system follows detailed balance and Markovianity, it can be shown that these eigenfunctions form a complete orthonormal basis i.e, $\langle \psi_i|\psi_j\rangle _\pi=\delta_{ij}$ with a bounded eigenvalue spectrum $1=\lambda_0>\lambda_1 \geq \lambda_2 \geq \dotsb$, where $\pi$ denotes the system's equilibrium probability. Thus the system's state $\chi_t(\textbf{x})$ at time $t$ can be represented as $\chi_t(\textbf{x}) = \sum_i \langle \psi_i |\chi_t \rangle_\pi \psi_i(\textbf{x})$ and its time evolution after time $k \tau$, where $k$ is an integer and $\tau$ represents the lag time, is given by,
\begin{equation}
\label{eq_TO}
\chi_{t+k\tau}(\textbf{x}) = \mathcal{T}^k \circ \chi_t(\textbf{x}) = \sum_i \langle \psi_i |\chi_t \rangle \pi \psi_i(\textbf{x})\text{exp}\left(\frac{-k\tau}{t_i}\right)
\end{equation}

Here $t_i = -\tau/\text{log}\lambda_i$ denotes the implied time scale of the eigenfunction $\psi_i$, and eigenvalues $\lambda_i$ represent the autocorrelation times. Thus, the system's behavior at long time scales can be deduced from slow eigenfunctions i.e., $\psi_i$'s  corresponding to large $\lambda_i$'s of the transfer operator $\mathcal{T}$.

A well-established technique for estimating the leading eigenfunctions of the transfer operator is the variational approach for conformation (VAC) dynamics \cite{noe2013variational,nuske2014variational,perez2013identification}. The key idea underlying VAC is to leverage the bounded and sorted nature of the eigenvalue spectrum of $\mathcal{T}$, and successively maximize $\tilde{\lambda}_i$ to approximate leading $\tilde{\psi}_i$ such that the condition, $\langle \tilde{\psi}_i|\tilde{\psi}_j \rangle _\pi = 0$ is satisfied. In this way, VAC is able to learn slow modes with the highest autocorrelation time.

Time-structure based independent component analysis \cite{noe2015kinetic,m2017tica} (TICA) is a particular implementation of VAC where RCs are constructed as the linear combinations of the input features or CVs. For example, Pande \textit{et al.} \cite{m2017tica} used TICA to learn a lower dimensional manifold and used it as the biasing variable for metadynamics and performed enhanced sampling. However, the use of linear transformations in the ANN model limits the expressivity of the RCs learned by TICA. In a later work, Pande \textit{et al.} introduced kernel TICA \cite{schwantes2015modeling} (kTICA) for learning RCs that are non-linear functions of the input features. Here, the main idea was the introduction of non-linearity by defining a kernel function for constructing an abstract feature space from input features using a pairwise similarity measure. However, later works \cite{chen2019nonlinear} identified several limitations of kTICA implementation involving high memory $O(N^2)$, and time complexity $O(N^3)$.

\begin{marginnote}
\entry{Transfer Operator ($\mathcal{T}_\tau$)}{Mathematically, $\mathcal{T}_\tau$ is a Perron-Frobenius operator at lag time $\tau$ that is the propagator of $u_t(\textbf{x})=\frac{p_t(\textbf{x})}{\pi(x)}$, where $\pi(\textbf{x})$ denotes equilibrium distribution function of the system's microstates, and $p_t(\textbf{x})$ represents the probability distribution of x at time t. Consequently, $u_{t+\tau}(\textbf{x}) = \mathcal{T}_\tau \circ u_t(\textbf{x})$.}
\end{marginnote}

One of the earliest applications of artificial neural networks (ANNs) in implementing the VAC principle for dimensionality reduction was the VAMPnets \cite{mardt2018vampnets}. The key idea in VAMPnets is the automated construction of MSMs from structural features where the ANN's loss function, called the VAMP-score, is derived using a Koopman operator framework. The authors noted one particular choice of VAMP-scores, the VAMP-2 is well suited for time series data which intuitively corresponds to the sum of the squared eigenvalues of the transfer operator \cite{wu2020variational}. However, the output layer of VAMPnets i.e, the reduced dimensions are constructed by applying a softmax function for detecting discrete metastable states of a system, and are thus not directly suitable for performing enhanced sampling which often requires smoothly differentiable variables. To utilize the VAMPnets framework for constructing continuous and descriptive RCs, Ferguson \cite{chen2019nonlinear} \textit{et al.} proposed the state-free reversible VAMPnets (SRV) for approximating the eigenfunctions of the transfer operator as nonlinear mappings of the input space without using the kernel trick. SRV is implemented through an ANN that maps input coordinates to an $n$-dimensional continuous output. Here, $n$ is a hyperparameter of the method representing the number of slow modes learned by the ANN model. In practice, the outputs $f_i(\textbf{x})$ of the ANN approximate the eigenfunctions as a linear combination $\tilde{\psi}_i(\textbf{x}) = \sum_js_{ij}f_j(\textbf{x})$ by minimizing, $\mathcal{L}=\sum_ig(\tilde{\lambda}_i)$. Here, $g(\tilde{\lambda}_i)$ is any monotonically decreasing function of the machine-learned eigenvalues of the transfer operator, $\tilde{\lambda}_i$. However, the authors noted that specific choices e.g, $g(\tilde{\lambda})=-\tilde{\lambda}^2$ or, $g(\tilde{\lambda})=1/log(\tilde{\lambda})$ where the loss function corresponds to maximizing the cumulative kinetic variance (VAMP-2 score) or, the sum of the implied time scales respectively produce good results.

\begin{marginnote}
\entry{Koopman operator}{Mathematical operator describing dynamical evolution of nonlinear functions under a linear framework. While transfer operators are concerned with predicting probability densities, Koopman operators focus on predicting functional values with time.}
\end{marginnote}

In a very recent publication, Ferguson \textit{et al.} \cite{shmilovich2023girsanov} extended SRV by proposing Girsanov Reweighting Enhanced Sampling Technique (GREST) which utilizes both dynamical Girsanov reweighting and thermodynamic corrections to learn the slow modes. As discussed in previous sections, a general scheme for RC construction involves iterations between performing MD simulations and analyzing biased simulation data. Naturally, this analysis will produce more accurate results when considering the biased nature of the accelerated MD trajectories and implementing dynamical corrections. Previously proposed approaches have visited this same problem in the context of the RAVE method \cite{wang2020understanding} by using the square-root formalism originally attributed to Bicout and Szabo \cite{bicout1998electron}. GREST solves the problem by modifying the SRV loss function according to the Girsanov theorem. The implemented ANN learns from a set of discontinuous, biased MD trajectories under simulation potential, $V_{sim}(\textbf{x})=V_{target}(\textbf{x})-U_{bias}(\textbf{x})$. The key idea behind Girsanov reweighting is to correctly assign path probabilities of different trajectories evolving under $V_{sim}(\textbf{x})$ to follow unbiased $V_{target}(\textbf{x})$. Finally, dynamical observables obtained from the path ensemble are used to estimate unbiased averages.

In a typical dimensionality reduction through enhanced sampling protocol, a crucial step in learning improved RCs through iterations is the initial first round of MD. The learned RC from the initial round can be improved by performing separate unbiased simulations where the system is initialized at different metastable states that are known \textit{a priori}. However, when studying complicated practical systems, it is possible that this will still not result in good initial RCs and it will require many iterations to get a converged result due to the absence of transition dynamics. Here, one strategy could be to perform several biased simulations e.g., using metadynamics by employing different trial RCs and attempting to combine information from the resultant trajectories. Even if the trial RCs are suboptimal, the learned improved RC from combining trajectories will likely be superior due to better sampling. However, combining biased trajectories with different RCs is not trivial and Parrinello \cite{bonati2021deep} \textit{et al.}, introduced Deep-TICA which is essentially a complete protocol for implementing this idea. The main idea behind Deep-TICA is to learn RCs from the first round of biased simulation by implementing a nonlinear VAC principle and biasing the leading eigenfunctions. It should be noted that different biased simulations will have different eigenvalue spectrums of the transfer operator as the modes that are accelerated will be different across simulations. In Deep-TICA this is taken into account by using the accelerated time scale \cite{tiwary2013metadynamics} for implementing the VAC principle and learning improved RCs. Additionally, Deep-TICA employs the recently developed OPES multithermal for the first round of biased simulations using the potential energy of the system as a RC. The subsequent rounds of simulations are also implemented through OPES due to its rapid convergence to a quasistatic regime.
 
\subsection{INFORMATION BOTTLENECK-BASED APPROACHES}
\label{subsec:inf_bottle}
\begin{marginnote}
\entry{Information bottleneck (IB)}{Information-theoretic framework aimed at constructing a faithful, yet simplified  representation of input data that retains the most relevant information for making accurate predictions.}
\end{marginnote}

Information bottleneck \cite{tishby2000information} (IB) based approaches perform dimensionality reduction of MD data by optimizing the trade-off between complexity and prediction accuracy of an ML model. The resulting low-dimensional manifold, called the latent space can be used as the biasing RC of an appropriate enhanced sampling scheme such as umbrella sampling or metadynamics. Typical implementations of IBs involve autoencoder \cite{hinton2006reducing} type ML models that consist of two sequential feed-forward ANNs known as the encoder and decoder respectively. During training, the encoder parameters are optimized to generate a latent space embedding containing the most essential information in the training data while the decoder parameters are optimized to reconstruct the input from the latent space as accurately as possible. Thus, the encoder-decoder pair is trained simultaneously to find an optimal trade-off in a self-supervised manner. A key advantage of IBs in contrast to the transfer operator approximation methods discussed earlier (see Section \ref{subsec:transfer_operator}), is that autoencoders consist of decoders that can effectively map data points sampled from the latent space back to the original input space, enabling them to function as generative models.

\begin{marginnote}
\entry{Generative models}{ML algorithms that learn the underlying probability distribution of a dataset to generate new samples resembling the original data.}
\end{marginnote}

One of the earliest adoptions of autoencoders for MD was the Molecular Enhanced Sampling with Autoencoders (MESA) by Ferguson \cite{chen2018molecular} \textit{et al.}. MESA is used to learn a non-linear mapping between atomic coordinates ($\textbf{x}_t$) generated from MD and a low dimensional latent representation that can reconstruct $\hat{\textbf{x}}_t$ with minimal error, $\mathcal{L} = \sum_{q=1}^Q\| \textbf{x}_{q,t}-\hat{\textbf{x}}_{q,t}\|^2$. The optimal dimensionality of the autoencoder is determined by the fraction of variance explained by the latent space compared to the input data. Since, MESA works with atomic coordinates, translational and rotational motion is taken into account through mean centering and augmenting MD data with rotationally invariant MD frames respectively. Finally, the learned RCs are used to perform umbrella sampling (US) and an estimate of the unbiased free energy surface of the MESA RCs is computed using the Weighted histogram analysis method \cite{kumar1992weighted} (WHAM).

In contemporary work, Wehmeyer and Noe \cite{wehmeyer2018time} proposed the time-lagged autoencoder (deep TAE) where input coordinates ($\textbf{x}_{t}$) at time $t$ is given to the encoder and the decoder learns to predict atomic coordinates ($\textbf{x}_{t+\tau}$) at a later time ($t+\tau$). The authors showed that linear TAEs correspond to time-lagged canonical correlation analysis (TCCA) and are equivalent to TICA if the MD data is time-reversible. By performing non-linear transformations and choosing appropriate lag-time $\tau$, highly expressive RCs for enhanced sampling can be obtained.

\begin{marginnote}
\entry{Posterior distribution}{Probability distribution over the latent space given the observed data. In a generative model, new data samples can be generated by sampling the learned posterior distribution.}
\entry{Evidence lower bound (ELBO)}{In variational Bayesian techniques, ELBO provides an estimate that is lower than or equal to the logarithm of the likelihood of observed data.}
\end{marginnote}

Pande \textit{et al.} extended this approach of considering time-lag in reconstruction error computation and proposed variational dynamics encoder \cite{hernandez2018variational} (VDE) which substitutes vanilla autoencoders with variational autoencoders \cite{kingma2013auto} (VAEs). In VAEs, the latent space is assumed to follow a prior distribution, typically a multivariate Gaussian distribution and the encoder learns to approximate the posterior distribution through training. During training, VAEs aim to maximize the evidence lower bound \cite{blei2017variational} (ELBO) consisting of two terms: the reconstruction loss ($\mathcal{L}_R$) measuring how well the VAE can reconstruct the input data, and the Kullback-Leibler (KL) divergence ($\mathcal{L}_{KL}$) between the approximate posterior and the prior distribution. The KL divergence encourages the approximate posterior to match the prior distribution, promoting regularization and controlling the complexity of the latent space. Under the VDE framework a third term, the autocorrelation loss ($\mathcal{L}_{AC}$) is also considered which maximizes the largest dynamical eigenvalue by following the VAC principle and encourages the discovery of the slowest process in the input data. Thus, the final VDE loss function takes the form, $\mathcal{L} = \mathcal{L}_R+\mathcal{L}_{KL}+\mathcal{L}_{AC}$.

As discussed in previous sections, the key strategy in computing improved RCs for enhanced sampling is to iterate between simulations and data analysis. However, after conducting the first round of the biased simulation, the effect of the deposited bias on MD trajectory should be taken into account for the accurate identification of RCs in subsequent rounds. Tiwary \textit{et al.} addressed this issue by proposing the Reweighted Autoencoder for Variational Bayes \cite{ribeiro2018reweighted} (RAVE). RAVE redefines the reconstruction loss by taking the deposited metadynamics bias ($V_{bias}$) into account, $\mathcal{L}_R=\sum_iw_i^2(\textbf{x}_{i,t}-\hat{\textbf{x}}_{i,t})^2$, where $w=e^{V_{bias}/k_bT}$. Unlike the previously discussed IB-based approaches so far the encoder of RAVE adopts a linear activation function to keep the latent space interpretable. It should be noted that, unlike deep-TAE or VDE, the original version of the RAVE decoder does not implement any time lag and aims to reconstruct the input data. However, in a later work \cite{wang2019past}, the authors introduced time-lag by modifying the RAVE objective function to $\mathcal{L} = I(\chi,\textbf{X}_{\Delta t}) - \beta I (\textbf{X},\chi)$. Here, the first term is the mutual information between latent representation $\chi$ at time $t$, and input ($\textbf{X}_{\Delta t}$) at a later time $t+\Delta t$. $I(\textbf{X},\chi)$ represents the mutual information between input and latent representation at time $t$. The hyperparameter $\beta$ can be used to tune the trade-off between model complexity and prediction accuracy. While implementing this objective function the authors reported improved performance when setting $\beta=0$ as it reduces the number of model parameters and adds a stochastic term to the decoder reconstruction for avoiding data memorization. Wang \textit{et al.} further improved the RAVE protocol \cite{wang2020understanding} for iterative biasing by correcting the RAVE objective function to take into account the effect of bias for small time-lag on the dynamical propagator of the system.

In another modified and as of now the preferred version of RAVE, Wang \textit{et al.} introduced state predictive information bottleneck \cite{wang2021state,mehdi2022accelerating} (SPIB) which aims to learn the metastable states of a system in addition to learning a low-dimensional representation of MD data given a time delay $\Delta t$. SPIB acts as a fast mode filter that ignores fluctuations occurring at a time scale smaller than the time delay when generating the latent space suitable for enhanced sampling. From \textit{a priori} system information, a user provides guessed initial metastable state assignment and the stochastic, non-linear decoder iteratively reconstructs and refines the metastable state definitions until convergence. Additionally, SPIB implements a mixture of Gaussians known as VampPrior \cite{tomczak2018vae} as the prior distribution instead of a single Gaussian for improved regularization of the latent space.

\subsection{EIGENDECOMPOSITION TECHNIQUES}
\label{subsec:spectral_decom}

In this section, we look at data-driven methods for RC discovery that aim to find a low-dimensional representation of MD trajectories by computing the eigendecomposition of relevant operators other than the transfer operator discussed earlier (Section \ref{subsec:transfer_operator}).

\begin{marginnote}
\entry{Path collective variables (path CVs)}{Class of path-like CVs that aim to describe transitions between predefined states present in a molecular system.}
\end{marginnote}

An interesting approach for enhanced sampling was proposed by Parrinello \textit{et al.} \cite{rizzi2019blind} that involves linear discriminant analysis (LDA). LDA is a supervised ML algorithm that constructs a ($d-1$) dimensional linear projection ($\textbf{W}$) of the input data with $d$ labeled classes. This is achieved by maximizing the inter-class distance ($\textbf{S}_b$) and minimizing the within-class variance ($\textbf{S}_w$) simultaneously, which is equivalent to maximizing the Fisher's ratio: $\underset{\textbf{W}}{argmax} \frac{\textbf{W}S_b\textbf{W}^T}{\textbf{W}S_w\textbf{W}^T}$. Using this expression it is trivial to show that the eigenvector corresponding to the largest eigenvalue ($\lambda$) of the generalized eigenvalue problem ($S_b\textbf{W}=\lambda S_w \textbf{W}$) serves as the LDA projection operator. For data with two labeled classes (A \& B) with given expectations ($\bm{\mu}$) and variances ($\bm{\sum}$), $\textbf{W}$ can be computed using $\textbf{W}=(\bm{\sum}_A + \bm{\sum}_B)^{-1}(\bm{\mu}_A - \bm{\mu}_B)$ \cite{duda2006pattern}. In the context of MD, the expectations and variances of the system CVs can be obtained from independent, short unbiased simulations initialized from states A \& B that are separated by a high energy barrier. The authors noted that this definition of projection operator assigns CVs with high variance higher importance, causing poor sampling of the more stable CVs with small fluctuations. However, the space spanned by the latter type of path CVs needs to be adequately explored to overcome the high energy barrier. To address this issue, the authors proposed taking the harmonic average of the variance instead of the arithmetic average when constructing the projection operator. The modified scheme is termed harmonic linear discriminant analysis (HLDA) and using Eq. \ref{eq_hlda}, HLDA can construct path CVs without any \textit{a priori} path information. Here the subscript denotes the class for which the expectation or variance has been calculated.

\begin{equation}
\label{eq_hlda}
     \textbf{W}=\frac{1}{\frac{1}{\bm{\sum}_A} + \frac{1}{\bm{\sum}_B}}(\bm{\mu}_A - \bm{\mu}_B)
\end{equation}

In a later work, Bonati \textit{et al.} \cite{bonati2020data} extended this approach by proposing Deep-LDA. Here, input CVs recorded from short unbiased simulations initialized from two separate states are passed through a non-linear feed-forward ANN. The output of the last hidden layer acts as a high-level feature identifier and is used as the input to an ANN implementation of LDA. Here the ML model is trained to maximize the largest eigenvalue of the generalized eigenvalue problem mentioned previously to ensure maximal class separation. It should be noted that deep-LDA employs a traditional definition of the projection operator involving arithmetic average instead of the definition introduced in HLDA.

 Very recently, Hocky \textit{et al.} \cite{sasmal2023reaction} proposed a novel LDA implementation for identifying optimal path CVs by taking position coordinates of relevant atoms as inputs instead of CVs. This framework is particularly useful for studying systems where the number of input CVs scales poorly with system size (e.g, pairwise distance scales quadratically). To directly work with atomic coordinates as they scale linearly, the authors reformulated the generalized eigenvalue problem to a generalized singular value decomposition (SVD). This enables the consideration of singular $S_w$ matrices for the computation of the projection operator which is often observed for position coordinates. After removing the effects of translational and rotational motion represented in position coordinates by aligning the molecular structures to a global average, optimal path CVs are computed and biased using OPES.

\begin{marginnote}
\entry{Principle of Maximum Caliber}{Concept in statistical mechanics suggesting that the most unbiased probability distribution is the one that maximizes entropy while satisfying all known constraints such as average values of certain variables, conservation laws, or other known properties of the system.}
\end{marginnote}

Compared to the LDA based methods discussed above, Tiwary and Berne took an alternative approach by proposing the spectral gap optimization of order parameters (SGOOP) method \cite{tiwary2016spectral}. The key idea in SGOOP is that the transition probability matrix ($\bm{\Omega}$) corresponding to the low-dimensional RC will have the largest time-scale separation between its slow and fast modes. If $\lambda_0=1>\lambda_1\geq \lambda_2 ...$ represent the ordered eigenvalues of $\bm{\Omega}$, then this is achieved by finding parameters of the learned RC that describe it as a linear or non-linear combination of the input CVs to maximize the spectral gap i.e, $\lambda_s - \lambda_{s+1}$, where $s$ is the number of apparent barriers. Here, the principle of Maximum Caliber with constraints ($\rho_i$) is implemented for estimating the transition matrix from an ensemble of short, unbiased/biased simulations where bias would otherwise prevent recovery of kinetics. Here, $a$, and $b$ represent discretized state of the system in the learned low-dimensional space. In a later work \cite{smith2018multi}, Smith \textit{et al.} extended this approach by proposing iterative construction of the low-dimensional RC starting from 1-d by employing conditional probability factorization to focus on transitions that are not captured in earlier dimensions. Here, additional components of the RC are constructed only if the prior component failed to capture a slow mode. In a recent work \cite{tsai2021sgoop}, the authors introduced a stopping criterion for the iterative addition of orthogonal components to the RC by computing commute time on the space of a kinetically accurate distance measure termed SGOOP-d.

\subsection{OTHER DIMENSIONALITY REDUCTION APPROACHES}
\label{subsec:other_dim_red}

In this section, we report notable and recently developed machine-learned-dimensionality reduction techniques that do not fall under the classes of methods presented in previous sections.

Subsequent to Rydzewski \& Nowak introducing the t-Distributed Stochastic Neighbor Embedding (t-SNE) \cite{van2008visualizing} for representing MD data, Zhang \& Chen proposed a t-SNE based enhanced sampling scheme \cite{zhang2018unfolding} that generates a stochastic kinetic embedding of the input CVs into a low-dimensional RC. By assuming the input CVs undergo an implicit diffusion process, here the KL divergence between high-dimensional transition probability matrix ($\textbf{M}_{input}$) and low-dimensional $\textbf{M}_{t-SNE}$ from  is minimized through a multilayer perceptron. Finally, the exploration of the current least informative region in the CV space is performed through well-tempered metadynamics in an iterative manner. Rydzewski \& Valsson extended this approach through multiscale reweighted stochastic embedding (MRSE) \cite{rydzewski2021multiscale}, by incorporating a multiscale representation of the input CVs and thus removed the choice of perplexity, which is a hyperparamter of t-SNE from the protocol. t-SNE involves computing the probabilities $p_{ij}, q_{ij}$ of choosing a sample $\textbf{x}_i$ as a neighbor of the sample $\textbf{x}_j$ in the feature ($\textbf{M}$), and low-dimensional ($\textbf{Q}$) space respectively. In a traditional t-SNE implementation, perplexity controls the trade-off between the captured local and global properties which can be difficult to select without \textit{a priori} system information. Under this framework, the pairwise probability distribution $\textbf{M}_{mix}$ is constructed by considering a mixture of $\textbf{M}$'s stemming from individual perplexity values. An ML algorithm is implemented to parametrize model parameters that maximize the KL divergence between $\textbf{M}$, and $\textbf{Q}$ for a batchsize $N_b$. The final loss function is shown in eq. \ref{eq_t_SNE}. For an accurate construction of the RC, a landmark selection scheme is adopted to prevent the under-representation of transition state data points during training, and a reweighting scheme to learn from biased data are adopted.

\begin{equation}
\label{eq_t_SNE}
     \mathcal{L} = \frac{1}{N_b}\sum_{i=1}^{N_b}\sum_{i=1,i \neq j}^{N_b} p_{ij} \text{log}\frac{p_{ij}}{q_{ij}}
\end{equation}

An alternative approach \cite{sun2022multitask} to dimensionality reduction was adopted by Kozinsky \textit{et al.} which adopts a multitask learning scheme. This method modifies a VAE architecture by adding a second, parallel decoder layer that minimizes potential energy error for learning a more informed transition state in the latent space. Under this framework, the objective function of the ML model is a weighted sum of three loss functions, $\mathcal{L} = c_c \mathcal{L}_c + c_p \mathcal{L}_p + c_r \mathcal{L}_r$. Here, the subscripts $c,p,r$ denote metastable state classification, potential energy error minimization, and latent space regularization tasks respectively. It should be noted that the metastable state labels for $\mathcal{L}_c$ are obtained from a TPS scheme and through committor analysis which can become expensive with the presence of many metastable states.

\section{LEARNING NEW STRATEGIES TO ENHANCE SAMPLING}
\label{sec:LearningEnhance}
Although the majority of work incorporating ML into enhanced sampling has been on dimensionality reduction, a few creative methods have used ML to inform how new simulations are initialized or how bias is deposited. In the adaptive sampling community, this has primarily been achieved through reinforcement learning-inspired methods that define a goal or reward function that is then used to initialize new simulations. These methods balance initialization in regions that optimize a desired property and initialization in undersampled regions that may contain pathways to optimal regions. In the biasing community, ML has been used to develop new strategies for bias deposition in contrast to the kernel density estimation used by metadynamics and OPES. These methods have primarily achieved this goal by introducing new neural network-based free energy estimators but have also treated bias deposition as a reinforcement learning problem.
\subsection{REINFORCEMENT LEARNING STRATEGIES FOR ADAPTIVE SAMPLING}
\label{subsec:AdaptiveSampling}

\begin{figure}
    \centering
    \includegraphics[
    width=\textwidth
    ]{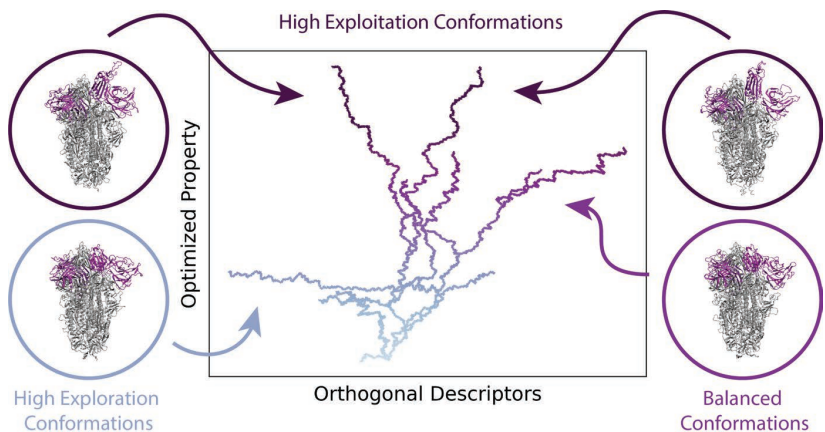}
    \caption{An illustrative example of the trade-off between exploration and exploitation in reinforcement learning. Here we show a collection of cartoon trajectories for conformational change in a protein. The optimized property is the distance to a reference conformation where distant conformations are shown in blue and close conformations are shown in purple. Reinforcement learning methods balance sampling regions such as the bottom left conformer which is in a poorly explored region of configuration space (exploration) and sampling regions such as the top conformers which have already optimized the property of interest (exploitation).}
    \label{fig:reinforcement}
\end{figure}

Three methods, FAST \cite{FAST}, REAP \cite{REAP}, and AdaptiveBandit \cite{AdaptiveBandit}, take inspiration from reinforcement learning to treat adaptive sampling initialization as a policy selection problem \cite{bandit_problems, ExploreExploit, RLAlgorithms, RLSurvey} where decisions are made to maximize a reward function.

FAST \cite{FAST} takes inspiration from the multi-armed bandit problem \cite{bandit_problems} to initialize simulations that are likely to optimize a given property such as root-mean-squared deviation (RMSD) to a target state. The multi-armed bandit problem weighs the trade-off between exploration and exploitation when facing uncertainty in rewards. In the original problem, a gambler faces the trade-off between exploiting a slot machine (also called a one-armed bandit) with known rewards or exploring for a potentially better machine. FAST, however, balances the trade-off between sampling states with optimal values of the metric of interest (exploitation) and poorly sampled states (exploration) by seeding simulations in proportion with a reward function that balances these two. An illustrative example of the general trade-off for reinforcement learning adaptive sampling is shown in \textbf{Figure \ref{fig:reinforcement}}. The reward $r(c_i)$ for state/cluster $c_i$ shown in Eq. \ref{eq:FAST} combines a directed component $\bar{\phi}(c_i)$ that rewards states with a higher value of the metric of interest and an undirected component $\bar{\psi}(c_i)$ that rewards poorly sampled states with $\alpha$ controlling the weight of exploration and exploitation.

\begin{equation}
    \label{eq:FAST}
    r(c_i) = \bar{\phi}(c_i)  + \alpha \bar{\psi}(c_i)
\end{equation}

REAP \cite{REAP} uses a reinforcement learning-inspired reward function to select starting configurations in the setting where a set of CVs is known but their importance is not. REAP calculates a reward function for each cluster $c_i$ shown in Eq. \ref{eq:REAP} in order to seed simulations in clusters with high rewards. The reward function standardizes each CV $\Theta_j$ and then takes a weighted sum of their absolute values. The standardization is done with respect to the mean and standard deviation of all clusters $C$ and the weights $w_j$ ranging from 0 to 1 are interpreted as the importance of $\Theta_j$. The weights are updated iteratively after each round of sampling to maximize the reward over the set of least sampled clusters. This iterative scheme allows sampling to proceed in different directions as new states are discovered.

\begin{equation}
    \label{eq:REAP}
    r(c_i) = \sum_{j = 1}^{k} w_j \frac{\lvert \Theta_j(c_i) - \langle \Theta_j(C)\rangle \rvert}{\sigma_j(C)}
\end{equation}

AdaptiveBandit \cite{AdaptiveBandit} is another method based on the multi-armed bandit problem but in this setting the goal is to find minimum free energy configurations. The reward is defined as the negative mean of the free energy of configurations sampled after a starting point and simulations are initialized using the UCB1 algorithm \cite{UCB1} for the multi-armed bandit problem. UCB1, shown in Eq. \ref{eq:UCB1}, defines a trade-off between the expected reward $Q_t$ for a given action $a$, or initialization in this case, and the uncertainty based on how many times the action has been selected. The uncertainty is defined as the square root of the ratio of the total past actions taken $t$ and the number of times action $a$ has been taken $N_t(a)$. AdaptiveBandit uses an MSM to discretize the system's states and estimate their free energy, providing a discrete set of choices for the starting state and estimated rewards. For each iteration, a random configuration is chosen from the optimal starting state according to UCB1, a short trajectory is sampled, then the MSM is updated which updates the possible choices, the associated rewards, and their uncertainties.

\begin{equation}
    \label{eq:UCB1}
    a_t = \underset{a}{\mathrm{argmax}}\left[ Q_t(a) + c \sqrt{\frac{\mathrm{ln}(t)}{N_t(a)}}\right]
\end{equation}

\subsection{LEARNING NEW STRATEGIES TO BIAS SIMULATIONS}
\label{sec:biasing}

\begin{figure}
    \centering
    \includegraphics[
    width=\textwidth
    ]{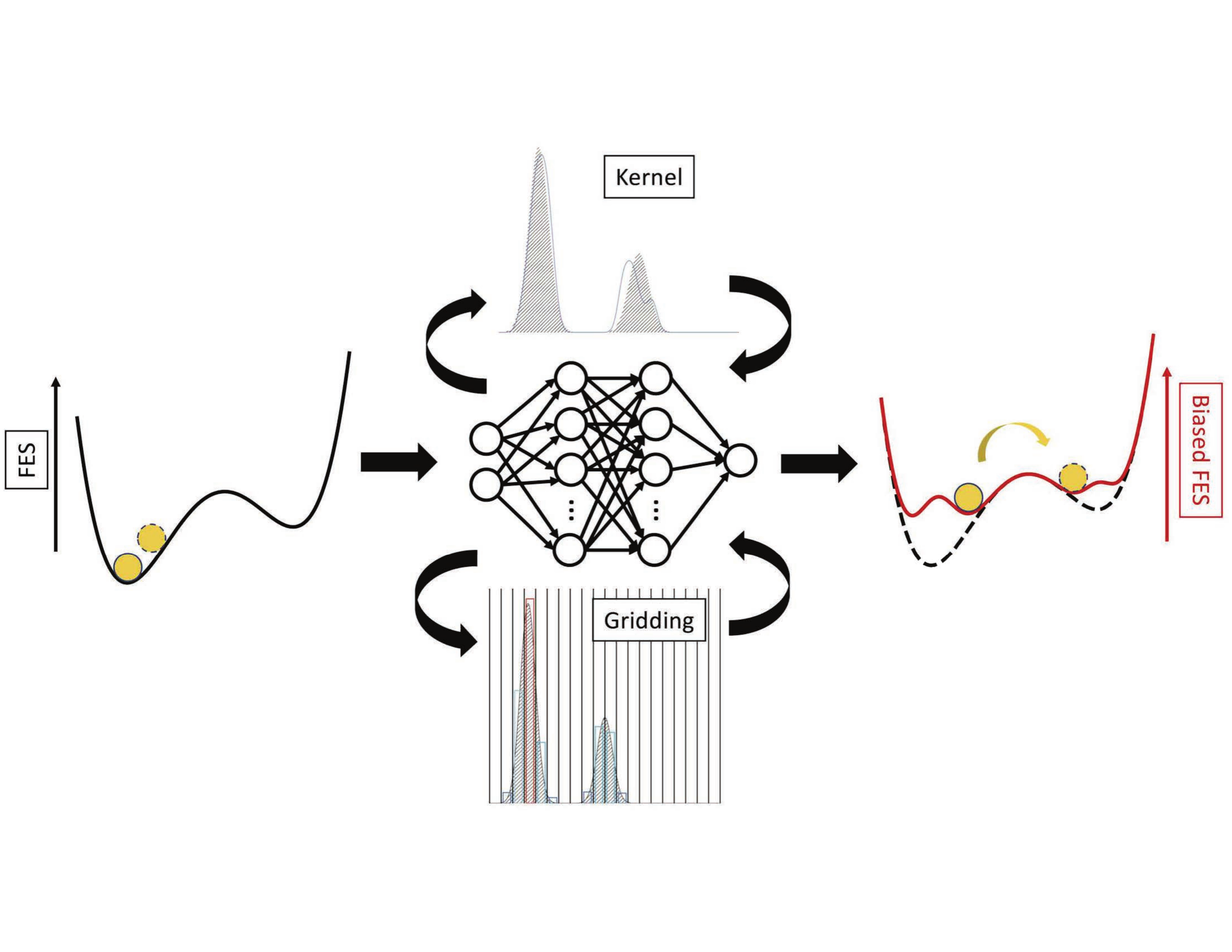}
    \caption{Schematic of machine-learned biasing methods reviewed in Section \ref{sec:biasing}, where typical estimators (Gaussian kernel density estimation (top) \cite{scikit-learn} and discrete grids (bottom)) are replaced with trained ANN/DNN models to yield high accuracy and efficient computation on the biases. }
    \label{fig:biasing}
\end{figure}

In this section, we review free energy-based biasing methods in which the process of modifying the Hamiltonian through the deposition of bias or through other approaches is improved by ML algorithms (see \textbf{Figure \ref{fig:biasing}} for an illustration). Both non-parametric methods and artificial neural networks (ANNs) have been directly integrated with enhanced sampling methods to facilitate sampling efficiency and accuracy. While there are several examples of such integration, here we highlight three.

Cs\'{a}nyi and collaborators introduced Gaussian Process Regression (GPR), a non-parametric ML technique to reconstruct functions in multiple dimensions, on the reconstructed free energy surface first from the umbrella sampling method \cite{stecher2014free}, and later on to enhance the sampling of a combination of adaptive biasing force and metadynamics methods \cite{mones2016exploration}. Specifically, metadynamics is used to deposit bias potentials, and instantaneous collective forces (ICF) are estimated using the adaptive biasing force method for the final reconstruction of the final multidimensional free energy surface. It has been shown that with this metadynamics/ICF/GPR scheme, the sampling/computational efficiency is significantly improved.

Besides combining the two enhanced sampling methods metadynamics and adaptive biasing force with the kernel-based ML method reviewed above, ANNs, which require training on prebuilt datasets, have also been applied individually to these two methods. As the second example we mention work by Galvelis and Sugita \cite{galvelis2017neural}. Here an ANN is trained on estimating free energy on-the-fly for an input higher-dimensional CV space. Such an approximated free energy is then used to construct a bias potential before additional biasing with traditional CV-based metadynamics. This method in principle not only allows fast approximations of the instantaneous free energy but also a higher-dimensional biasing scheme. However, as pointed out by the authors \cite{galvelis2017neural}, limitations show up when applied to more complicated systems with high-dimensional biases being deposited. In such systems, the construction of bias potential becomes less efficient and eventually makes it difficult for transitions to occur with any computational advantage. Later on, this very similar strategy was applied to another free energy based biasing method, called the variationally enhanced sampling (VES) method \cite{valsson2014variational}. In particular, the high-order basis set expansion needed in VES is replaced by such ANN that takes pre-screened CVs as input and predicts free energies of the CV space. With this approach, the computational cost becomes manageable and the number of input CVs is no longer limited (the amount of variational parameters scales exponentially with the number of chosen CVs in conventional VES) \cite{bonati2019neural}.

As the third example, we describe how ANNs were combined with the adaptive biasing force method. In the traditional approach, discrete grids are used to estimate the mean forces and consequently the free energies. However, with such a scheme, a tradeoff between precision and efficiency needs to be made: fast convergence can be reached under low-resolution grids, but numerical issues arise on the estimates of mean force as regions become broader. In the ``Force-biasing Using Neural Networks" (FUNN) approach \cite{guo2018adaptive}, a self-regularizing ANN \cite{sidky2018} is trained to provide an on-the-fly estimate on free energy and, later on, to learn the generalized mean force. Specifically, the model is optimized with rigid regularization to make the network robust to hyperparameters and overfitting. Results show that both ANN sampling techniques speed up the sampling compared to their non-NN counterparts, and additionally, FUNN leads to faster convergence regarding ANN sampling.

The methods we highlighted above mix machine-learning techniques with traditional adaptive biasing methods --- metadynamics and adaptive biasing force methods. Naturally, this recipe can be used to create more flavors of data-driven biased simulations. As an example, the idea of reinforcement learning has been introduced to enhance the sampling of unexplored regions along selected CV spaces in \cite{zhang2018reinforced}. In a similar spirit as metadynamics, Gaussian biases are deposited at regions where sufficiently sampled in the reinforced dynamics method. Unlike metadynamics, the bias deposition in \cite{zhang2018reinforced} relies on an uncertainty indicator, defined as the standard deviation of the predictions outputted from the trained deep neural network models. A significant advantage of such a formalism is that it allows one to enhance the sampling in high-dimensional CV spaces, given the ability of deep neural networks in representing high-dimensional functions. However, as also acknowledged by the authors, the quality of selected CVs affects the resulting approximation of free energy surface and approaches to construct ML RCs are reviewed in Section \ref{subsec:CV_sel}. Recently, deep learning methods were also applied to Gaussian accelerated Molecular Dynamics (GaMD). Unlike the typical GaMD method \cite{miao2015gaussian}, Do \textit{et al.} \cite{do2023deep} introduce machine-learned boost potentials to smoothen the system's potential energy surface, termed Deep Boosted Molecular Dynamics (DBMD). These potentials are optimized by iterative training on the randomly generated boost potentials in initial GaMD runs and a chosen anharmonicity threshold is used as a sign of convergence. We believe we will be seeing many more such methods in the coming years mixing different ML architectures with different enhanced sampling protocols for data-driven biased simulations that in principle do not require dimensionality reduction at least on the part of the user.


\section{ESTIMATING FREE ENERGIES WITH FLOW-BASED MODELS}
\label{sec:flow}

\begin{marginnote}
\entry{Flow-Based Model}{Deep learning methods which learn to transform a complex probability distribution $p(\mathbf{x})$ to a simple probability distribution $q(\mathbf{x})$ with the same dimensionality of $p(\mathbf{x})$.}
\end{marginnote}

In contrast to dimensionality reduction methods, flow-based models do not attempt to map the dynamics of the system to a simpler, lower-dimensional manifold. The key idea is to instead transform the complex probability distribution of the data into a more tractable distribution while preserving the dimensionality of the data. As such, flow-based models are especially suited for the study of systems with highly complex structures and dynamics; systems with many metastable states and no clear separation of timescales may not be accurately represented on low-dimensional manifolds. Flow-based models learn the distribution of states in the full configuration space, avoiding dimensionality reduction and assumptions about the data made therein.

\begin{marginnote}
\entry{Generative Modelling}{A class of ML techniques which attempts to generate realistic samples resembling an empirical data distribution.}
\end{marginnote}

More precisely, flow-based models formulate generative modeling as learning deterministic or stochastic mappings from a complicated empirical distribution to a simple prior distribution. Once the map bridging the empirical and prior distributions is learned, samples from the prior can be transformed into samples that resemble the empirical data distribution, typically at a low computational cost. Since flow-based models learn mappings between probability distributions, they have been primarily used to estimate free energies as opposed to learning dynamics, which remains an open, exciting area.

\begin{figure}
    \centering
    \includegraphics[
    width=\textwidth
    ]{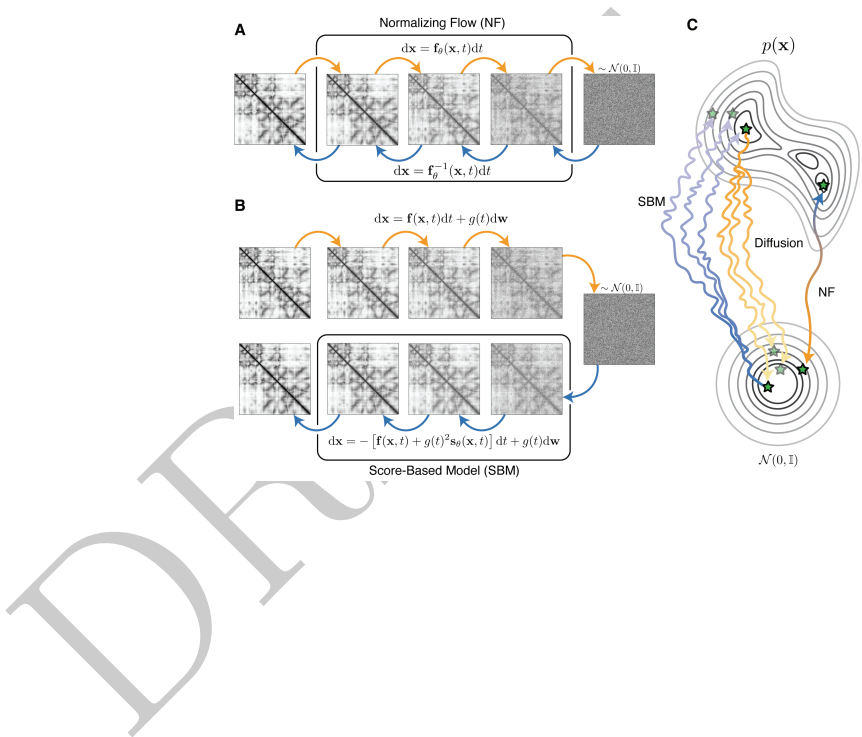}
    \caption{\textbf{A} Depiction of the learning process in normalizing flows, where samples from a complex data distribution are mapped to a simple prior distribution. The mappings in both directions are learned simultaneously. \textbf{B} Illustrative depiction of how a score-based model maps samples from a complex data distribution to a simple prior distribution. Notably, only the mapping from the prior distribution to the data distribution is learned. \textbf{C} Comparative illustration emphasizing the difference between score-based models and normalizing flows; score-based models learn only the reverse mapping, whereas normalizing flows parameterize the forward and reverse mappings simultaneously. And while normalizing flow mappings are bijective, mappings learned by score-based models will produce different results for each realization of the generative processes.}
    \label{fig:flow-fig}
\end{figure}

\subsection{NORMALIZING FLOWS}
\label{sec:norm-flow}

Normalizing flows are models that establish an invertible mapping between empirical and target distributions while ensuring that the Jacobian of the mapping remains computationally tractable \cite{nf-review}. These models have been employed in the enhanced sampling community to improve the accuracy of free energy estimation and enable the efficient generation of a large number of realistic samples.

\begin{marginnote}
\entry{Normalizing Flows}{A class of flow-based models which learn a transformation between prior $q$ and data-distribution $p$, which is exactly invertible and has a tractable Jacobian.}
\entry{Likelihood Maximization}{Approach to inference which maximizes the probability of known data being generated by the inference model.}

\entry{Jacobian}{A matrix, $J(f, \mathbf{x})$, whose entries describe the rate of change of the dimensions of a vector-valued function with one another. Mathematically, the entries take the form $J_{ij} = \frac{\partial f_i}{\partial x_j}$, where the subscripts denote the dimension of vector $x$ and $f$ evaluated over $x$.}

\entry{Partition Function}{The normalization constant associated with a distribution, typically denoted as $Z$. Free energy estimation amounts to estimating $\log Z$ for configurations of a physical system.}
\end{marginnote}

Normalizing flows parameterize an invertible mapping $f_\theta:\mathrm{\mathbf{x}}\rightarrow \mathrm{\mathbf{z}}$ between data $\mathrm{\mathbf{x}}$ sampled from an empirical distribution to $\mathrm{\mathbf{z}}$ that follows a target or prior distribution that can be easily sampled. \cite{neural-ode, real-nvp, normalizing-flow-variational}. Such a mapping may be parameterized as an ordinary differential equation taking the form
\begin{equation}
    \label{eq:flow-ODE}
    \frac{\mathrm{d\mathbf{x}}}{\mathrm{d}t} = f_\theta(\mathrm{\mathbf{x}},t)
\end{equation}
with learnable drift $f_\theta$ and solution
\begin{equation}
    \label{eq:flow-ODE-solution}
    \mathrm{\mathbf{z}} = \mathrm{\mathbf{x}} + \int_0^1 f_\theta(\mathrm{\mathbf{x}}(t))\mathrm{d}t.
\end{equation}
Eq. \ref{eq:flow-ODE-solution} transforms densities of $\mathrm{\mathbf{x}}$ to those of $\mathrm{\mathbf{z}}$ by integrating the drift $f_\theta(\mathrm{\mathbf{x}}(t))$ over the interval $[0,1]$. Since $f_\theta$ is invertible, the reverse mapping amounts to replacing the drift with $f_\theta^{-1}$ and interchanging $\mathrm{\mathbf{x}}$ and $\mathrm{\mathbf{z}}$ in Eq. \ref{eq:flow-ODE-solution}. See \textbf{Figure \ref{fig:flow-fig}A} for a schematic representation of the mapping learned by normalizing flows.

The mappings set forth in Eqs. \ref{eq:flow-ODE} and \ref{eq:flow-ODE-solution} are appealing since they admit exact likelihood maximization under certain normalization conditions. In particular, the density of $p(\mathrm{\mathbf{x}})$ under the image of an invertible mapping $f$ is 
\begin{equation}
    \label{eq:flow-reweighting}
    p(f(\mathrm{\mathbf{x}})) =\frac{p(\mathrm{\mathbf{x}})}{|\det J_f(\mathrm{\mathbf{x}})|},
\end{equation}
 where $J$ is the Jacobian of $f$. The significance of Eq. \ref{eq:flow-reweighting} becomes clear when one recognizes that $|\det J_f(\mathrm{\mathbf{x}})|$ is the factor by which the partition function, $Z(\mathbf{x})$, changes under $f$. The partition function of a complicated distribution can be estimated by establishing an invertible mapping to a distribution with a tractable partition function and evaluating Eq. \ref{eq:flow-reweighting} for $f^{-1}$.

Na\"ive efforts to take advantage of Eq. \ref{eq:flow-reweighting} may be thwarted by the computational demand of evaluating $|\det J_f(\mathrm{\mathbf{x}})|$. The normalizing flow architectures address the computational complexity associated with evaluating the Jacobian by imposing restrictions on the operations performed by the network. An elegant example is the RealNVP architecture \cite{real-nvp}, which utilizes layers that result in upper-triangular or lower-triangular Jacobians. In this case, the determinant simplifies to the product of the diagonal elements. By alternating between upper and lower-triangular operations, the architecture enables efficient computation of determinants while maintaining universal approximation.

However, this efficiency comes at a cost. While these restrictions make normalizing flows computationally tractable, they limit their expressivity and make it challenging to learn complex mappings between distributions. Normalizing flows have been observed to struggle with mapping multi-modal empirical distributions to unimodal priors \cite{BBVI}. Recent advancements in normalizing flows aim to address these issues. For example, one approach involves alternating between deterministic flow layers and Monte-Carlo sampling layers \cite{SNF, NF_MCMC_1, NF_MCMC_2}. This approach is similar to the class of generative models known as score-based generative models (discussed in Section \ref{sec:scorebased}).

\subsection{BOLTZMANN GENERATORS}
\label{sec:boltzmann-gen}
The pioneering use of normalizing flows to enhance sampling is the Boltzmann generator \cite{noe2019boltzmann}. The Boltzmann generator establishes an invertible mapping between the distribution of system conformations $\mathbf{x}$ and a latent distribution $\mathbf{z}$ using a normalizing flow. Once this mapping is learned, it becomes possible to transform samples from the prior distribution into realistic conformations through the inverse mapping.

\begin{marginnote}
\entry{Boltzmann Distribution}{The distribution corresponding to thermodynamic equilibrium, which is of the form $p(x) = e^{-\beta u(\mathbf{x})}/Z$, where $\beta$ is inverse temperature, $u(\mathbf{x})$ is the potential energy of conformation $\mathbf{x}$, and $Z$ is the partition function.}
\end{marginnote}

However, it should be noted that the generated conformations are not guaranteed to follow a Boltzmann distribution, so thermodynamic free energies cannot be computed directly. To account for this, the Boltzmann generator employs a reweighting scheme based on the potential energies of the conformations, $u(\mathbf{x})$, in which the probabilities are rescaled by the Boltzmann factor $e^{-\beta u(\mathbf{x})}$. The reweighting scheme utilized by the Boltzmann generator is limited to implicitly solvated systems because it requires generating \textit{all} degrees of freedom of the system. For explicitly solvated systems, it would be necessary to generate the degrees of freedom associated with the solvent, which becomes computationally infeasible even for relatively small systems.

Other variants of the Boltzmann generator attempt to integrate with multi-ensemble simulation protocols, incorporate transferability across thermodynamic parameters, and improve the robustness of the learned normalizing flow \cite{LREX, smooth_normalizing_flow, SNF}.

\subsection{FREE ENERGY ESTIMATION THROUGH INVERTIBLE MAPPINGS}
\label{sec:invertible-mappings}

\begin{marginnote}
\entry{Configuration space}{For an $N$-particle system, the configuration space is the $3N$ dimensional space in which points represent particular configurations or arrangements of the $N$-particle system.}

\entry{Multistate Bennet Acceptance Ratio}{A widely used data-driven approach to estimate free energies, abbreviated as MBAR, which performs maximum likelihood inference over data generated under different thermodynamic conditions (e.g. temperatures), making MBAR especially useful in conjunction with exchange-based simulation protocols.}
\end{marginnote}

Following the Boltzmann generator, there has been a proliferation of approaches that combine normalizing flows with analytically informed approaches to free energy estimation. The accuracy of free energy estimates depends on the degree of configuration space overlap between the states being compared. Unfortunately, unbiased simulations often exhibit limited overlap between states, leading to slow convergence. Methods like BAR (Bennett Acceptance Ratio) and MBAR (Multistate Bennett Acceptance Ratio) aim to improve estimates by explicitly maximizing the likelihood of the free energy \cite{BAR, MBAR}. Notably, MBAR has been shown to be a statistically optimal free energy estimator for uncorrelated samples.

Another approach to enhancing the convergence of free energy estimates is finding an invertible mapping $\mathcal{M}$ which increases the overlap between states in configuration space and has a tractable Jacobian \cite{Jarzynski_2002_TFEP}. Normalizing flows are particularly suitable for this task. Consequently, there has been a line of research focused on learning normalizing flow mappings $f_\theta$ which increase overlaps in configuration space and maximize the likelihood of the free energy \cite{Rizzi_TFEP, Wirnsberger_2020_TFEP, nf-atomic-solids}. The free energies under the transformation $f_\theta$ can then be reweighted using Eq. \ref{eq:flow-reweighting} to obtain an estimate of the conformational free energy.

\subsection{SCORE-BASED MODELS}
\label{sec:scorebased}
 Just as normalizability is fundamental to normalizing flows, stochasticity plays a central role in score-based models. Score-based models extend the framework of normalizing flows to encompass stochastic differential equations (SDEs). These models encompass a variety of frameworks that approach generative modeling as learning stochastic processes \cite{ddpm, score-sde, score-langevin, stochastic-interpolants}. Drawing inspiration from non-equilibrium thermodynamics, score-based models capture complex relationships within data \cite{diffusion-noneq}.

Score-based models can be expressed as pairs of forward and backward SDEs:
\begin{equation}
\label{eq:sde-fwd}
\text{d}\mathbf{x} = -f(\mathbf{x},t)\text{d}t + g(t)\text{d}\mathbf{w}\quad \text{and} \\
\end{equation}
\begin{equation}
\label{eq:sde-bck}
     \text{d}\mathbf{x} = -\left[f(\mathbf{x},t)  + g(t)^2\nabla_{\mathbf{x}}\log p_t(\mathbf{x})\right]\text{d}t + g(t)\text{d}\mathbf{w}.
\end{equation}
Here, $f(\mathbf{x},t)$ represents the drift guiding the diffusion process and $g(t)$ parameterizes the time-dependent noise. The forward and reverse diffusions are described by Eq. \ref{eq:sde-fwd} and Eq. \ref{eq:sde-bck} respectively, and are evaluated from $t=0$ to $t=T$ for the forward process and $t=T$ to $t=0$ for the reverse. 

The forward diffusion, determined by choice of $f(\mathbf{x})$ and $g(t)$, maps an empirical distribution $p(\mathbf{x})$ to the stationary distribution of $f(\mathbf{x})$. When $f$ is linear in $\mathbf{x}$, the stationary distribution of Eq. \ref{eq:sde-fwd} is a Gaussian, allowing for efficient computation of the diffusion process at any given time $t$. Unlike normalizing flows, the forward process is not learned (see \textbf{Figure \ref{fig:flow-fig}B} and \ref{fig:flow-fig}C).

Evaluating the reverse process (Eq. \ref{eq:sde-bck}) is less straightforward due to the dependence of the term $\nabla_\mathbf{x}\log p_t(\mathbf{x})$ on the initial conditions of the forward diffusion. This term, referred to as the score $\mathbf{s}(\mathbf{x},t)$, is parameterized by a neural network. Intuitively, the score opposes the gradient of the probability flow in Eq. \ref{eq:sde-fwd}. Similar to normalizing flows, score-based models are trained by maximizing a variational lower bound on the score, often referred to as the score-matching objective. 

\begin{marginnote}
\entry{Score}{The gradient of the free energy (likelihood) of a stochastic process. The score indicates the direction of probability flow and is the central quantity estimated in score-based generative modeling. A stochastic differential equation can be approximately inverted by observing realizations, estimating the score, and then moving opposite the score.  }
\end{marginnote}

Score-based models also admit unbiased estimation of the Jacobian of $\mathbf{s}(\mathbf{x},t)$, which allows for reweighting of probability densities as in Eq. \ref{eq:flow-reweighting}. Similar to normalizing flows, the network used to parameterize the score can reinforce inductive biases of the data \cite{transformer-diffusion, RosettaDiffusion}. By drawing upon theories developed for analyzing stochastic processes, score-based models are an appealing inductive prior for applications of deep learning to molecular dynamics \cite{stochastic-proc-MD}.

\subsection{INTEGRATION WITH ENHANCED SAMPLING FRAMEWORKS}
\label{sec:DDPM}

Flow models -- deterministic or stochastic -- do not independently solve the sampling problem. Both classes of models have been coupled with existing statistical mechanics frameworks of sampling to enhance the sampling of thermodynamic observables. The most prominent examples of this approach couple normalizing flows with free energy perturbation and replica exchange methods \cite{LREX, DDPM-REMD, nf-atomic-solids}. 

In this review, we will specifically focus on the coupling approach employed by our group, which involves post-processing replica exchange simulations to enhance free energy estimation \cite{DDPM-REMD}. In replica exchange simulations, the rate of exchange scales as $1/\sqrt{N}$ for an $N$-particle system, which represents a crucial limitation of the method, especially for applications to large systems. Even with low replica exchange rates, diffusion models recover Boltzmann weights and provide accurate free energy estimates within the distribution of simulated temperatures. When extrapolating outside the simulated range, diffusion models exhibit greater robustness in estimating free energies compared to other data-driven methods such as MBAR.

As of the time of writing this review, the integration of score-based models with enhanced sampling frameworks remains an active area of research. In the machine learning community, score-based models have emerged as a more robust alternative to normalizing flows which share many attractive mathematical properties. While training normalizing flows on molecular systems can be an arduous task, requiring expert knowledge of the system to construct a suitable prior \cite{LREX, nf-atomic-solids}, score-based models have little trouble mapping highly complex dynamics to trivially simple priors. We anticipate that the robustness of score-based models, coupled with enhanced sampling frameworks, will facilitate the study of complex systems of great cross-disciplinary interest.





\section{DISCUSSION}
\label{sec_outlook}

Recent advances in ML augmented enhanced sampling techniques such as the ones we discussed in this Review have had a profound impact on the capabilities of MD simulations. These advances have expanded the scope of studying rare events and complex dynamical processes, leading to a deeper understanding of molecular behavior \cite{rizzi2021role,rogal2019pathCV,vani2023alphafold2,spiwok2022collective}. Despite these advances, there are still potential research areas that warrant further exploration.


\subsection{BENCHMARK APPLICATIONS}
\label{sec_benchmarks}
After the development of an enhanced sampling technique, the next crucial step is to assess its effectiveness by testing it on various systems. The methods discussed in this review so far typically examine metastable state transitions in analytical systems like the double/four-well, Müller–Brown potential, or small molecular systems such as alanine dipeptide, chignolin, Trp-cage, villin headpiece and others \cite{lindorff2011fast}. However, the lack of standardized benchmark systems presents challenges in accurately evaluating the performance of these methods. To address this, it would be beneficial to the enhanced sampling community to establish a curated set of test systems specifically designed to assess an ML model's robustness (e.g., to hyperparameter tuning), efficiency (training data requirements for enhanced sampling), and finally the ability of the method in overcoming different kinds of thermodynamic, kinetic bottlenecks. Additionally, new test systems should be regularly proposed to prevent methods from being developed solely to surpass the existing benchmarks, otherwise knowingly or unknowongly prior familiarity with the system can influence true assessment of the methods being developed. By adopting such standardized evaluation criteria, appropriate comparisons between new and existing methods can be made to assess their strengths and weaknesses in a transparent manner.

\subsection{MODEL INTERPRETABILITY \& EXPLAINABILITY}
\label{sec:interpretablility}
One of the key reasons behind increased adoption of ML techniques in MD can be attributed to their ability to learn from complicated data distributions in an automated, and efficient manner. However, highly expressive ML models typically come with the cost of poor interpretability i.e, it becomes difficult to understand why an ML model is making its predictions. This loss of interpretability makes identification of inaccurate ML models difficult, especially when the model starts memorizing training data by overfitting model parameters. Thus, it is worthwhile to examine and assign a degree of trust to a trained ML model prior to adopting it for enhanced sampling \cite{carvalho2019machine}.

This challenge can be addressed by either designing, (i) inherently interpretable ML models from MD data or, (ii) a post hoc interpretation scheme for explaining the behavior of complicated ML models. Methods such as RAVE \cite{ribeiro2018reweighted} takes the former approach by adopting a linear encoder for enhanced sampling, while deep-LDA \cite{bonati2020data} discussed in Section \ref{sec:dim_red} proposes using the modulus of the weights between input and first layer of the ANN for interpretation. However, most of the methods discussed in this review implement non-linear transformations to achieve high expressivity and as a result, the latter approach must be adopted to make the models explainable. To this end, Mehdi \textit{et al.} proposed Thermodynamically Explainable Representations of AI and other blackbox Paradigms (TERP) that constructs linear, interpretable surrogate models to approximate local behavior of the complicated ML model \cite{mehdi2022thermodynamics}. TERP is inspired by methods that are well-established in the general domain of ML, and has been designed to be suitable for MD data. The authors of this review hope to see more active investigations in the domain of intepreting ML models used for MD.
 
\subsection{LEARNING MEANINGFUL RCs}
\label{sec_learning_RCs}

The primary aim of dimensionality reduction techniques for enhanced sampling is the construction of informative RCs (Section \ref{sec:dim_red}). If the RCs are able to capture system behavior with sufficient detail, the learned ML model can even be implemented in similar but different systems through transfer learning eliminating the need for retraining \cite{sultan2018transferable}. Additionally, depending on the task at hand, the constructed RCs can be further improved by imposing constraints. For example, one promising approach could be to isolate different types of thermodynamic and kinetic bottlenecks along orthogonal components of the constructed RC and employ enhanced sampling techniques that are most suitable for overcoming the respective bottlenecks. In a recent work, Beyerle \textit{et al.} \cite{beyerle2022quantifying} employed SPIB \cite{wang2021state} to successfully learn disentangled energy-entropy coordinates in the machine-learned RC for certain model systems. In another work \cite{wang2022introducing}, Wang \textit{et al.} introduced Dynamics Constrained Auto-encoder (Dynamics-AE), which constructs a latent space that follows a prior probability distribution based on overdamped Langevin dynamics instead of a typical gaussian distribution for more faithful and disentangled representations of physical systems.

\subsection{EXPLOITING SYMMETRY THROUGH MACHINE LEARNING}
\label{sec_exploiting_symmetry}

As discussed in Section \ref{sec:dim_red}, implementing
dimensionality reduction techniques to learn system RCs for enhanced
sampling involves the analysis of CVs such as torsion angles, pairwise distances, etc. These traditional CVs have rotational and translational invariance because they use internal coordinates but using them forces models to rely on a hand-picked basis set. This limitation becomes evident when attempting to sample self-assembly or nucleation processes, where repeated molecules form the system of interest, and traditional CVs are inadequate
for an accurate description of the system. RCs describing the state of the system by averaging neighboring interactions have been developed \cite{steinhardt1983bond, ten1998computer,giberti2015insight, tsai2019reaction,zou2021toward,zou2023driving}, but learning new basis functions has remained difficult. Ideally, these RCs could be learned from all-atom coordinates using a neural network that explicitly preserves these symmetries such as a graph neural network (GNN). However, implementing such an approach is nontrivial,
and research in this area is still in its early stages. In a very recent study \cite{liu2023graphvampnets}, Huang \textit{et al.} proposed GraphVAMPnets employing GNNs, capable of capturing local atomic information to learn an RC that preserves translational and rotational invariance. In the future, we expect to see further research in this direction, particularly focusing
on enhancing the transferability of the ML models, since the treatment of symmetries and the construction of underlying graphs may vary. 

\subsection{ROBUST FREE ENERGY ESTIMATION}
The conformational free energy is a fundamental quantity in the molecular sciences that is often computationally intractable to estimate, due to its relationship to the partition function. In recent years, the emergence of normalizing flows has made conformational free energy estimation tractable for complex systems, although the robustness of flow-based estimates of the free energy remains a critical barrier to widespread use.

 Recently, score-based models have shown greater expressiveness and robustness compared to normalizing flows, particularly when learning mappings from simple to highly complex distributions \cite{score-sde}. The construction of a suitable prior for normalizing flows can be challenging and typically relies on expert knowledge of the system \cite{nf-atomic-solids, LREX}. In contrast, diffusion models can reliably learn mappings from simple priors to highly complex data distributions and do not suffer from mode-seeking behavior to the same extent as normalizing flows \cite{DDPM-REMD, BBVI}. In fact, there has been a recent focus on enhancing the stability and improving the accuracy of both normalizing flows and score-based models through the integration of Monte-Carlo or importance sampling into the generative process \cite{SNF, NF_importance_sampling, NF_MCMC_1, NF_MCMC_2, score-sde}. Due to these advantages, we anticipate that score-based modeling will emerge as a more robust alternative to normalizing flows for enhanced free energy estimation, while still offering the desirable properties of normalizability and invertibility.

\section*{DISCLOSURE STATEMENT}
\label{sec_disclosure}
The authors are not aware of any affiliations, memberships, funding, or financial holdings that
might be perceived as affecting the objectivity of this review.

\section*{ACKNOWLEDGMENTS}
\label{sec_acknowledgment}
S.M. thanks the NCI-UMD Partnership
for Integrative Cancer Research for financial support. Z.S., Z.Z. and L.H. were supported by the National Institute of General Medical Sciences of the National Institutes of Health under Award Number R35GM142719. The content is solely the responsibility of the authors and does not represent the official views of the National Institutes of Health. P.T. was an Alfred P. Sloan Foundation fellow during preparation of this manuscript. We thank Deepthought2, MARCC,and XSEDE (projects CHE180007P and CHE180027P) for computational resources.

\bibliographystyle{ar-style3}

\end{document}